\documentstyle [12pt] {article}

\catcode`@=12
\begin{document}
\vspace{0.5in}
\oddsidemargin -.375in
\newcount\sectionnumber
\sectionnumber=0
\def\be{\begin{equation}}
\def\ee{\end{equation}}
\begin{flushright} UH-511-806-94\\September 1994\
\end{flushright}
\vspace {.5in}
\begin{center}
{\Large\bf S-wave phase shift in $\Lambda$$\pi$ scattering \\}
\vspace{.5in}
{\bf Alakabha Datta} and
{\bf Sandip Pakvasa\\}
\vspace{.1in}
 {\it
Physics Department, University of Hawaii at Manoa, 2505 Correa
Road, Honolulu, HI 96822, USA.}\\
\vskip .5in
\end{center}

\vskip .1in
\begin{abstract}

We calculate the s-wave strong interaction $\Lambda$$\pi$ phase shift
 at the $\Xi$ mass taking
into account contributions from the ${{\frac{1}{2}}}^{-}$ and
 ${{\frac{3}{2}}}^{-}$ $\Sigma$ resonances. We find the S-wave phase shift to
be small, of the order of 0.3 degrees and bounded by 0.5 degrees.

\end{abstract}
\vskip .25in

Recently there has been a calculation of $\Lambda \pi$ strong
interaction phase shift using baryon chiral perturbation
theory where it was found that in the leading order the
s-wave phase shifts, taking into account the $1/2^+, 3/2^+$  $\Sigma$
states, vanish \cite {ml}.  Calculations of $\Lambda \pi$ phase shifts are
relevant to the measurement of CP violation in the hyperon
decay $\Xi \rightarrow \Lambda \pi$ \cite{pak}. An experiment to measure the
combined asymmetry $ \Delta\alpha = \Delta\alpha_{\Lambda} +
\Delta\alpha_{\Xi} $ will be carried out in the near future at Fermilab
\cite{fl}. Here for example; $ \Delta\alpha_{\Xi} = \alpha_{\Xi} +
\bar{\alpha_{\Xi}}$ and $ \Delta\alpha_{\Xi} $ is proportional to $
\tan(\delta_s-\delta_p) $.
    In this short note we calculate the S-wave phase
shifts  by considering the $1/2^- \Sigma (1750) $ and the $ 3/2^-
\Sigma (1670)$ resonances \cite{PDG}.  In this case the S-wave phase shifts do
not vanish in the leading order in baryon chiral
perturbation theory.

As in Ref. 1, we use a $SU(2)_L$$\times$$ SU(2)_R$ invariant chiral
Lagrangian to describe the interaction of baryons with
pions.  We write the interaction Lagrangian as \cite{me,mee},

\begin{equation}
L_{int} \ = g_ {\Sigma \Lambda} \ \bar{\Lambda} \ \gamma_.A^b_a
\ \Sigma_{cb} \epsilon^{ac} \ + g_{\Sigma^* \Lambda}
\bar{\Lambda} \ \gamma_5
\ A^b_c \ \Sigma_{bc}^* \epsilon^{ac} \ + h.c.
\end{equation}
where,
\begin{equation}
 \Sigma = \left [
\begin{array}{cc}
\Sigma^+ & \frac{\Sigma^0}{\sqrt{2}} \\ \frac{\Sigma^0}{\sqrt{2}}
  & \Sigma^-\end{array} \right ]  \nonumber\\
\end{equation}
is the $\frac{1}{2}^-$ isotriplet and
$\Sigma^{* \mu}$ is the ${\frac{3}{2}}^{-}$ isotriplet with the
same matrix structure as $\Sigma$.
\begin{equation}
(A^{\mu}){^{b}}_a   =  \frac{i}{2} \left [ \xi \partial^\mu \xi^+ -
\xi^+ \partial^\mu \xi \right ]   \nonumber \\
\end{equation}
 where $\xi  =  exp \left [ \frac{iM}{f} \right ]$ and the meson
matrix is
\begin{equation}
M  =  \left [ \begin{array}{cc}
\frac{\pi^0}{\sqrt{2}} & \pi^+ \\
\pi^- & -\frac{\pi^0}{\sqrt{2}} \end{array} \right ]\nonumber\\
\end{equation}
with $f \sim 132 $ MeV
being the pion decay constant

In the heavy baryon limit of the chiral Lagrangian the
contribution from the $3/2^- \Sigma$ resonance vanishes.  The
leading order contributions to the phase shifts in the heavy
baryon expansion is suppressed by  $\frac{m_
\Sigma-m_\Lambda}{m \Sigma + m \Lambda} \sim .2$ and its
higher powers.  Hence the only contribution to the S-wave
phase shift comes from $1/2^- \Sigma$ state.  Using the
heavy baryon limit to (1) we replace $\gamma_\mu \rightarrow
v_{\mu}$; the baryon velocity, which is conserved and the
baryon propagator $\frac{i}{\gamma.P_{B}- m_{B}} \rightarrow
\frac{i}{v.k}$ where $P_B \ = m_B v + k$ is the
 baryon momentum and $m_B$
 is the baryon mass. In Fig.1 we show the diagrams that contribute to the
S-wave phase shift.

We calculate the S-wave phase shift to be
\begin{equation}
\delta_s =  - \frac{1}{2 \pi}  \frac{g^2_ {\Sigma \Lambda}}
{f^2}
       \frac{E_\pi^2 (E_\pi^2 - m_\pi^2)^{1/2}
(m_\Sigma - m_\Lambda)}{[E_\pi^2 - (m_\Sigma - m_\Lambda)^2]}\\
\end{equation}
we note that the phase shift calculation from the original Lagrangian
in eqn.(1) would result is an expression for $\delta_s$ similar to that in
eqn.(3) with corrections suppressed by
$\frac{m_\Sigma - m_\Lambda}{m_\Sigma + m \Lambda}$.  To
calculate $g_ {\Sigma \Lambda}$ we use the decay of $\Sigma(1750)$ to
$\Lambda \pi$.  However the branching fraction is unknown
and the total width has a large uncertainty.

Taking the branching fraction as a free parameter we solve for
$g_{\Sigma \Lambda}$ from the equation.
\begin{equation}
x\Gamma = \frac{g^2}{4 \pi f^2} (m_\Sigma -
m_\Lambda)^2 p_f \left [ \frac{E_\Lambda}{m_ \Sigma} +
\frac{m_\Lambda}{m_ \Sigma} \right ]
\end{equation}
with $ E_{\Lambda} = \frac{{m_\Sigma}^2 + {m_\Lambda}^2 -
{m_\pi}^2}{2m_\Sigma} $,
 $p_f = \sqrt{E_\Lambda^2 - m_\Lambda^2}$ where $\Gamma$ is the total
 width and $x$ is the
branching fraction.

In Fig. 2 we show a plot of the S-wave phase shift in
degrees for the pion-energy $E_\pi = 200 MeV$, which is the energy of
the pion in the decay $\Xi \rightarrow \Lambda \pi$, for
different branching fraction. The measured decay width and the branching
fraction are  $ 60 MeV \le \Gamma \le 160 MeV $ and $ x \le .75 $ \cite{PDG}.
 We therefore plot three different curves with $\Gamma=60,90$
and $160$ MeV respectively.
  We find that for a maximum
branching fraction of 75\% the S-wave phase shift can be at most $\sim
.5 $ degrees. Hence in conclusion we have shown that the S-wave
phase shifts from the $\Sigma$ $ {1/2}^-$ and $\Sigma$ ${3/2}^-$ resonances
for $\Lambda \pi$ scattering are small. We confirm the conclusion of Lu et
al. (Ref. 1) that the results for the phase shifts make
$\tan(\delta_s-\delta_p)$ for  $\Xi \rightarrow \Lambda \pi$ very small. If
these results for phase shifts are valid then
 $\Delta\alpha_{\Lambda}$ will dominate the measurement of $\Delta\alpha$ in
the forthcoming experiment. With these values for the phase shifts
$\Delta\alpha_{\Xi}$ is about 10 times smaller than previous estimates (Ref. 2)
 which
relied on earlier calculations of phase shifts \cite{m,nk}.

       The $\Lambda$-$\pi$ phase shifts near the $\Xi$ mass can be
measured by analysing the final states in the decay mode
$\Xi\rightarrow \Lambda \pi e \nu$ as shown by Pais and Treiman long
ago \cite{pt}. This mode has an expected branching ratio of order $10^{-7}$
 and in an expected
sample of $10^9$  $\Xi$'s there may be about 100 or so
$\Xi\rightarrow \Lambda \pi e \nu$ events in the data sample and it may be
possible to attempt such an analysis.

{\bf Acknowledgement}
This work was supported in part  by US
D.O.E grant \# DE-FG 03-94ER40833.

\subsection{Figure Captions}
\begin{itemize}

\item {\bf Fig .1}: Feynman diagrams that contribute to the S-wave phase
shift.

\item {\bf Fig .2}: Plot of the phase shift versus the branching fraction
for three different values of the total decay width of $\Sigma (1750)$ at
a pion energy of $E_\pi=.2GeV$.
 \end{itemize}
\end{document}